# Electrical control grain dimensionality with multilevel magnetic anisotropy

Shengyao Li[1†], Sabpreet Bhatti[1†], Siew Lang Teo[2], Ming Lin[2], Xinyue Pan[3], Zherui Yang[1], Peng Song[4], Wanghao Tian[4], Xinyu He[1], Jianwei Chai[2], Xian Jun Loh[2], Qiang Zhu[2,5,6], S. N. Piramanayagam[1*], Xiao Renshaw Wang[1,4*]

[1] *Division of Physics and Applied Physics, School of Physical and Mathematical Sciences, Nanyang Technological University, 637371, Singapore*

[2] *Institute of Materials Research and Engineering (IMRE), Agency for Science, Technology and Research (A*STAR), 2 Fusionopolis Way, Innovis #08-03, Singapore 138634, Singapore*

[3] *Cavendish Laboratory, University of Cambridge, 19 JJ Thomson Avenue Cambridge, CB3 0HE, Cambridgeshire, United Kingdom*

[4] *School of Electrical and Electronic Engineering, Nanyang Technological University, 50 Nanyang Ave, 639798, Singapore*

[5] *Institute of Sustainability for Chemicals, Energy and Environment (ISCE2), Agency for Science, Technology and Research (A*STAR), 1 Pesek Road, Jurong Island, Singapore 627833, Singapore*

[6] *School of Chemistry, Chemical Engineering and Biotechnology, Nanyang Technological University, 21 Nanyang Link, Singapore 637371, Singapore*

[†]These authors contributed equally

*Email: renshaw@ntu.edu.sg, prem@ntu.edu.sg

**Abstract**

In alignment with the increasing demand for larger storage capacity and longer data retention, electrical control of magnetic anisotropy has been a research focus in the realm of spintronics. Typically, magnetic anisotropy is determined by grain dimensionality, which is set during the fabrication of magnetic thin films. Despite the intrinsic correlation between magnetic anisotropy and grain dimensionality, there is a lack of experimental evidence for electrically controlling grain dimensionality, thereby impeding the efficiency of magnetic anisotropy modulation. Here, we demonstrate an electric field control of grain dimensionality and prove it as the active mechanism for tuning interfacial magnetism. The reduction in grain dimensionality is associated with a transition from ferromagnetic to superparamagnetic behavior. We achieve a non-volatile and reversible modulation of the coercivity in both the ferromagnetic and superparamagnetic regimes. Subsequent electrical and elemental analysis confirms the variation in grain dimensionality upon the application of gate voltages, revealing a transition from a multidomain to a single-domain state accompanied by a reduction in grain dimensionality. Furthermore, we exploit the influence of grain dimensionality on domain wall motion, extending its applicability to multilevel magnetic memory and synaptic devices. Our results provide a strategy for tuning interfacial magnetism through grain size engineering for advancements in high-performance spintronics.

**Table of content**

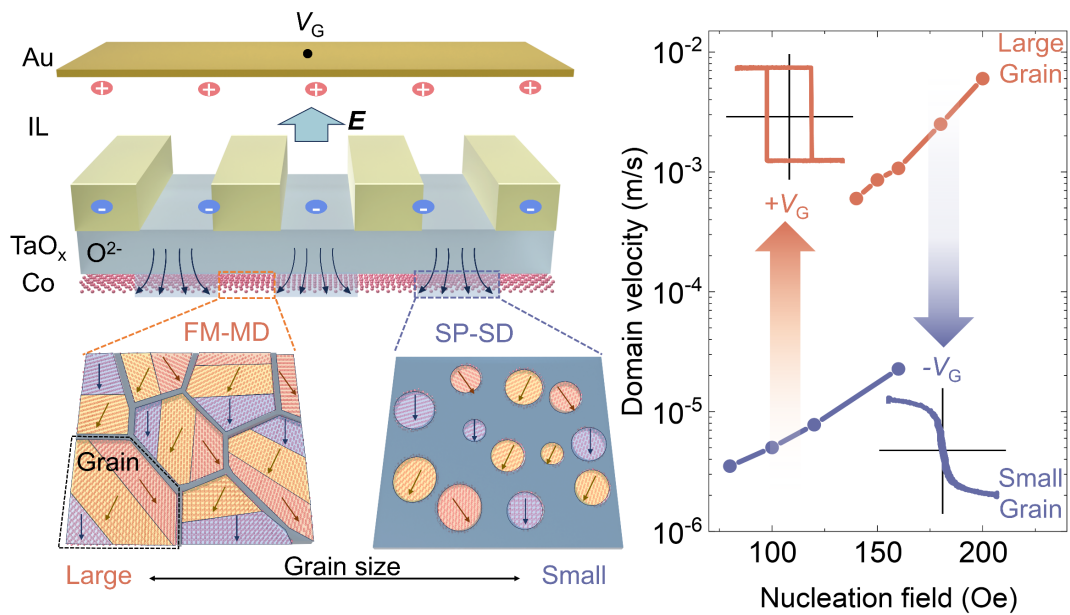

**Introduction**

Spintronics represents a cutting-edge field in modern electronics, with significant achievements in magnetic non-volatile memory technology.[1,2] A pivotal focus is to obtain diverse magnetic properties for various applications, e.g., the stability of magnetic bits depends on magnetic anisotropy energy, which is governed by the size of magnetic elements.[3,4] Magnetic properties are strongly correlated with the distribution of magnetic domains, determined by grain dimensionality and the interaction between grains.[4–6] Figure 1a schematically illustrates the evolution of grain dimensionality in a magnetic thin film, where grain dimensionality refers to the shape and size of individual grains. Large grains spontaneously divide into multiple magnetic domains, with the magnetization aligning uniformly within each domain. A reduction in grain size transforms the domain distribution from a multidomain (MD) to a single domain (SD) regime, leading to a magnetism transition from a ferromagnetic (FM) to a superparamagnetic (SP) state.[4,5,7] In this process, the grain dimensionality evolves from large planar grains to nanoparticles, separated by non-magnetic grain boundaries. Conventionally, grain size was controlled through various preparation techniques, including controlled film growth,[8,9] heat treatments,[10,11] and the utilization of seed layers.[12,13] Nevertheless, the grain dimensionality and magnetic properties obtained through these techniques are fixed upon fabrication and are not tunable at the device level.

Electric field control of magnetic properties is of particular interest in the field of spintronics. Due to its rapid response and compatibility with existing electronic systems, electrical manipulation of magnetic properties holds substantial promise for large-scale memory devices.[14,15] A typical device configuration involves a dielectric layer positioned between a gate electrode and a ferromagnetic layer,[16] thereby modulating the carrier density and subsequently altering magnetic properties under gate voltages ($V_G$).[15] In light of the emphasis on low-energy data storage, electrochemical gating control of interfacial magnetism has become a focal point.[17–21] This technique incorporates an electrolyte layer that functions as an ion reservoir. Upon the application of $V_G$, the electric field drives the ion migration, contributing to the magnetism modulation through the formation of compounds.[22–24] Numerous studies highlight the significant role of ion intercalation in magnetism modulation. However, few investigations explore its effects on grain dimensionality. Given the close relationship between grain dimensionality and magnetic properties,[25–27] understanding the influence of electric field-driven ion intercalation on grain dimensionality holds great potential for enhancing the modulation efficiency.

Here, we report the control of magnetic anisotropy through electrically tunable grain dimensionality. Adjacent to the magnetic layer, a $TaO_x$ layer serves as an $O^{2-}$ reservoir. Application of an external electric field induces $O^{2-}$ ion migration, resulting in the progressive oxidation of grain boundaries.[28] Consequently, this process leads to the enlargement and reduction of grain sizes. This ion intercalation induces a transition from a ferromagnetic state in a multidomain (FM-MD) regime, to a superparamagnetic state in a single-domain (SP-SD) regime. Notably, the coercivity ($H_c$) modulation in both regimes exhibits non-volatile and reversible characteristics. Both electrical and elemental analysis corroborate the reduction in grain dimensionality upon ion intercalation. The first-order reversal curves (FORC) reveal a decrease in exchange coupling, and cross-sectional elemental mappings indicate a diffusion of $O^{2-}$ ions. Furthermore, the electric field-controlled small grains exert a pinning effect on domain wall motion, which is extended to current-driven multilevel magnetization switching devices.

**Results and discussion**

$H_c$ reflects the ability of a ferromagnetic film to withstand an external magnetic field, which is intricately correlated with grain dimensionality. Figure 1b illustrates this relationship by displaying the variation of $H_c$ as a function of the grain dimension of our deposited film. All data points were obtained from anomalous Hall effect (AHE) measurements in Figure S21 of the Supporting Information. Under $V_G$ = -2 V, the ion migration leads to a decrease in grain size (accompanied by a reduced intergranular interaction), $H_c$ initially increases to a maximum value at a diameter of $D_s$, at which a conversion from MD to SD state occurs, then gradually decreases to zero. MD predominates for grains with a diameter larger than $D_s$, where larger grains subdivide into multiple domains. Below $D_s$, when the grain size becomes smaller, the anisotropy energy of the grains is reduced, resulting in random switching of the magnetization due to thermal energy. In the case of extremely small grains, $H_c$ diminishes to zero, signifying the superparamagnetic regime.

The influence of grain size on magnetic properties inspires us to design a device based on the deposited film that electrically controls the grain sizes, as depicted in Figure 1c. A detailed cross-sectional view of the device is presented in the zoom-in image. The device exhibits a multilayer structure composed of Ta(1 nm)/Pt(2 nm)/Co(0.8 nm)TaO$_x$(1.5 nm). Applying a negative $V_G$ induces the migration of anions to the interface between the electrolyte and the channel. At the same time, cations migrate to the interface between the electrolyte and the gate electrode. The presence of oppositely charged layers at the interface generates an electric double layer (EDL), thus establishing a strong electric field.[29–31] This electric field is highly effective in promoting the ionic migration within the TaO$_x$ layer, where $O^{2-}$ ions are inserted into the Co layer, leading to pronounced changes in electronic and magnetic properties.

Figure 1d depicts the evolution of AHE loops upon the application of $V_G$ for various durations, corresponding to the variation of grain size in Figure 1b. The as-deposited film shows a square hysteresis loop with a lower coercivity, marked by the red star, representing a ferromagnetic large multidomain (FM-MDL) state. In this state, neighboring planar grains exhibit significant intergranular interaction, leading to a rapid magnetization reversal. With the application of $V_G$ = -2 V, the intercalation of $O^{2-}$ ions segregates the grains and reduces the intergranular interaction. $H_c$

increases to the maximum at $D_s$, representing a ferromagnetic small multidomain (FM-MDS) state marked by the orange square. A longer duration of $V_G$ reduces $H_c$, marked by the blue triangle, representing the FM-SD state, where magnetic grains are isolated by non-magnetic boundaries. Finally, the purple sphere represents the SP-SD state, characterized by magnetic grains shrinking into isolated small particles with zero $H_c$. The result indicates that the application of $V_G$ is a promising way in tuning $H_c$ through the modulation of grain dimensionality.

The transfer curves and the leakage curves under various $V_G$ sweeping rates are provided in Figure S11 of the Supporting Information, while the area of transfer and leakage curves are shown in Figure 1e. The loop area of both curves gradually increases with a decrease in sweeping rate, indicating the intercalation of more ions by the extended $V_G$ sweeping process. Using the integrated charge method,[32] we can quantitatively estimate the amount of ions intercalated into the film to induce a modulation in magnetic anisotropy (Figure S12, Supporting Information).

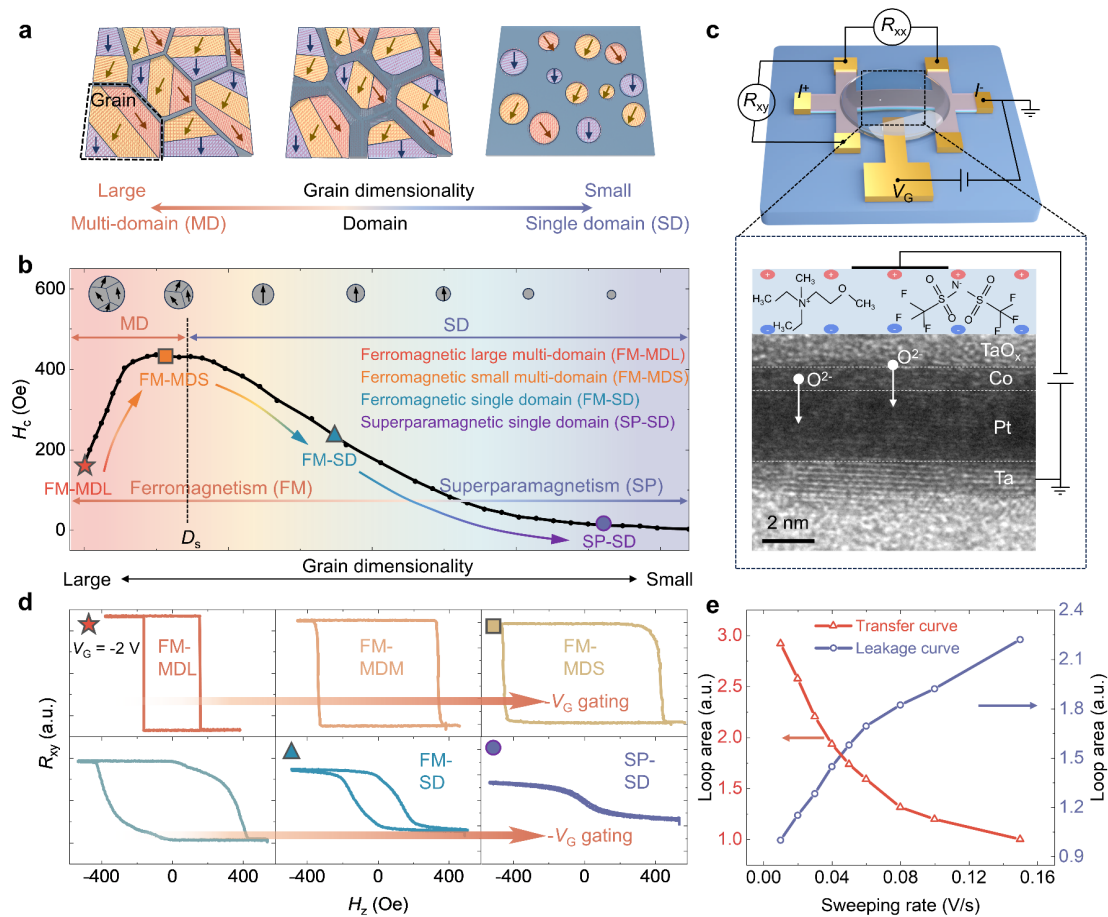

Figure 1. Magnetic property variation as a function of grain dimensionality. (a) Schematic illustration of magnetic domain distribution in grains of various dimensionality. (b) $H_c$ variation as a function of grain dimensionality. (c) Schematic image of the electrochemical gating device. The zoom-in displays the multilayer structure of a cross-sectional TEM image. The scale bar represents a length of 2 nm. (d) Representative AHE loops with various grain sizes upon the continuous

application of negative $V_G$. (e) The area of transfer and leakage curves as a function of $V_G$ sweeping rate.

Various ions have been investigated in controlling magnetism.[33–35] Here, we select $TaO_x$ as an $O^{2-}$ reservoir, which effectively modulates the Co grain size. Compared with other works, a thin layer of $TaO_x$ offers advantages of: (i) High oxygen mobility; (ii) Preservation of the magnetic properties of the Co thin film and enhancement in the perpendicular magnetic anisotropy (PMA); (iii) Facilitation of forming compounds with Co; (iv) Protection of the ultra-thin Co layer; (v) Ensuring high electric field from the EDL due to low resistivity of $TaO_x$. This $TaO_x$ layer adjacent to the Co layer significantly influences the structure of grain boundaries. Unlike well-crystallized grains, grain boundaries tend to undergo oxidation and alloying with neighboring elements.[28,36] The ultra-thin nature of the Co layer increases the likelihood of forming an oxygen-rich Ta-Co-Pt alloy at the grain boundaries, thereby isolating the Co grains.

Figure 2a displays a schematic representation illustrating the change in grain size and grain boundaries between FM-MDL and FM-MDS states. The intercalation of $O^{2-}$ leads to the formation of oxide-based grain boundary and a reduction in grain size, whereas the extraction of $O^{2-}$ reverses this process. Figure 2b displays the $H_c$ variation as a function of gating duration within the MD regime. In the left panel, $H_c$ increases under negative $V_G$ due to $O^{2-}$ intercalation, signifying that the enhanced grain boundary segregation leads to a reduction in the intergranular exchange coupling in the MD regime. Under a positive $V_G$, $O^{2-}$ extraction reverses FM-MDS to FM-MDL state (right panel). The details of the evolution of AHE loops are provided in Figure S18 of the Supporting Information. To demonstrate the reliability of this modulation, we repetitively switched $H_c$ for ten cycles between FM-MDL and FM-MDS states. A negative $V_G$ brings the $H_c$ of the FM-MDS state, while a positive $V_G$ restores the $H_c$ of the FM-MDL state as shown in Figure 2c. The right panel of Figure 2c exhibits the representative AHE loops at FM-MDL and FM-MDS states.

This electrically controlled ion migration also significantly impacts the grain size variation in the SD regime as schematically illustrated in Figure 2d. The intercalation of $O^{2-}$ further reduces grain size and probably the anisotropy constant by percolating into the grains, thereby resulting in the isolated small magnetic particles in an SP-SD state. Figure 2e presents the $H_c$ variation as a function of gating duration in the SD regime. In the right panel under negative $V_G$, $H_c$ initially experiences a rapid drop, eventually coming to zero at the SP-SD state, which indicates the gradual saturation of ion intercalation. In the right panel under positive $V_G$, $H_c$ exhibits a similar trend, rapidly rising initially and slowly back to the FM-SD state for a prolonged gating duration. Figure 2f shows the repeatable $H_c$ modulation between FM-SD and SP-SD states, and their corresponding AHE loops. Employing a negative (positive) $V_G$ brings the device from FM-SD to SP-SD (SP-SD to FM-SD) state. The alternation of grain size between FM-SD and SP-SD states exhibits reliable reversibility, with negligible degradation after functioning for hundreds of cycles (Figure S20, Supporting Information).

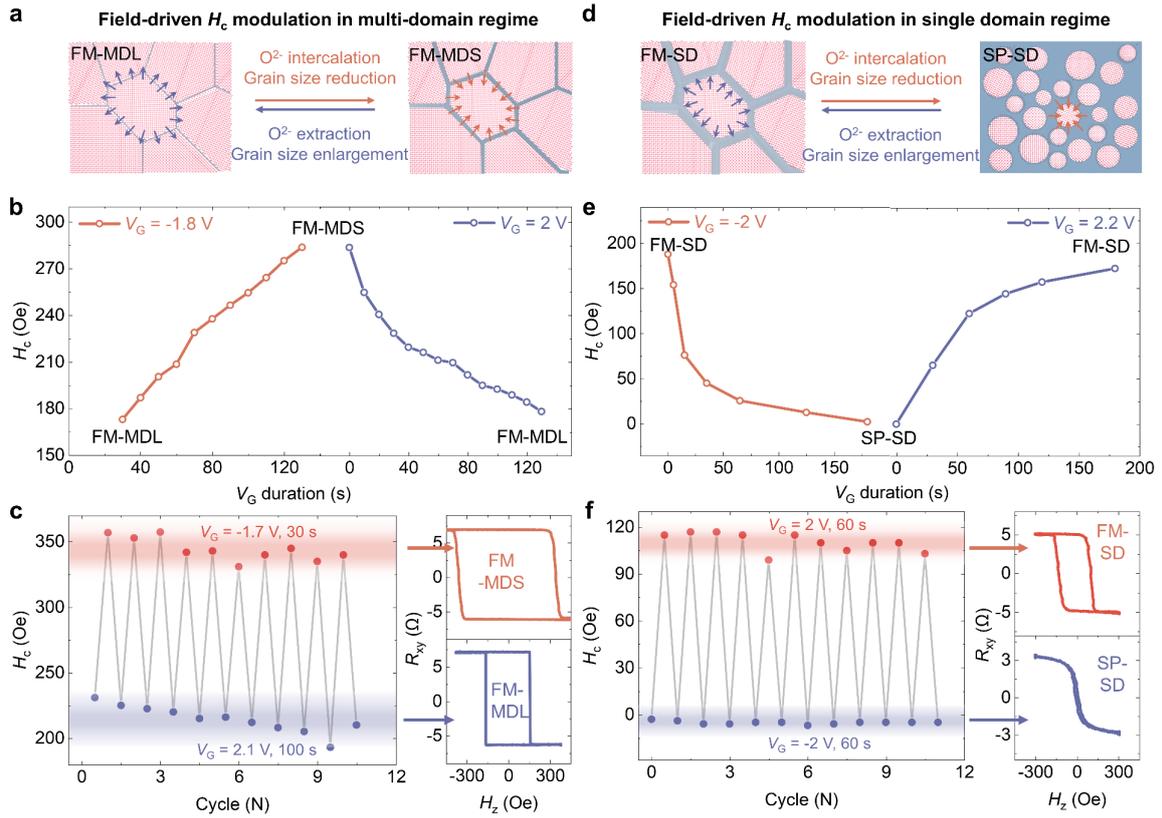

Figure 2. Multilevel, non-volatile, and reversible modulation of $H_c$. (a) Schematic of grain size modulation between FM-MDL and FM-MDS state. (b) $H_c$ modulation as a function of gating duration between FM-MDL to FM-MDS states. (c) Repetitive $H_c$ modulation between FM-MDL and FM-MDS states. The right panel displays the representative AHE loop in two states, respectively. (d) Schematic of grain size modulation between FM-SD and SP-SD state. (e) $H_c$ modulation as a function of gating duration between FM-SD and SP-SD states. (f) Repetitive $H_c$ modulation between FM-SD and SP-SD states. The right panel displays the representative AHE loop at FM-SD and SP-SD state, respectively.

To figure out the impact of this $V_G$-controlled ion migration on grain size and intergranular exchange interactions, we employ the FORC analysis. The FORC technique provides a comprehensive understanding of the switching fields and the interaction strength between grains, thereby effectively characterizing the transition between MD and SD.[38,39] The FORC measurements are proceeded by initially saturating the magnetization by applying a large magnetic field (higher than the saturation field $H_s$) and subsequently sweeping the field to a reversal field, $H_r$. The magnetic moment as a function of the field ($H$) is measured as the field swept back to $H_s$. By repeating the procedure for various $H_r$ values, a series of FORCs are obtained, as depicted in the inset of Figure 3a-c. The FORC distribution $\rho(H, H_r)$ is a mixed second derivative, given as

$$\rho(H, H_r) = \frac{-\partial^2 M(H, H_r)}{\partial H \partial H_r}. \tag{1}$$

Figure 3a-c show the FORC diagram of the device at FM-MDL, FM-MDS and FM-SD state, respectively. New coordinates can be built using

$$H_u = \frac{-(H + H_r)}{2}, H_c = \frac{H - H_r}{2}, \tag{2}$$

where $H_u$ represents the distribution of the interaction fields, and $H_c$ represents the distribution of the coercive field. The progression from FM-MDL to FM-SD state signifies an evolution of magnetic properties corresponding to increasing oxygen concentration. Notably, the contour plot exhibits a shift along $H_c$ and $H_u$ axes, with the intensity at the FM-MDL state being two orders of magnitude higher than that of the FM-SD state. This significant reduction in peak intensity reflects a decrease in exchange coupling.[8] The peak positions along the $H_c$ axis at three states are 210, 530, and 280 Oe, respectively. This trend aligns with the observed evolution of $H_c$ upon gating. Moreover, the elongation of the FORC contour along the $H_c$ axis implies the multiple magnetization reversal within the film, indicating the existence of multiple grains.[40] Therefore, the FORC results provide evidence of a reduction in grain size, which leads to the isolation of the magnetic domain following the electrochemical gating process.

The ion intercalation upon electrochemical gating leads to an elemental redistribution. Figure 3d presents the cross-sectional electron energy-loss spectroscopy (EELS) elemental distribution of the device at different states. At the FM-MDL state in the left panel, the full width at half maximum (FWHM) value of Co is 1.3 nm. The interface between $TaO_x$ and Co layers distinctly separates Co and O elements. Conversely, the FWHM of the Co element at the FM-SD state is 1.9 nm, as shown in the right panel, indicating the diffusion of Co. The elemental mapping indicates that both O and Co distributions at the FM-SD state are broader than that at FM-MDL state, making the interface less distinct. This elemental redistribution applies to all regions after electrochemical gating, as shown in Figure S22 of the Supporting Information.

The temperature-dependent $H_c$ also reveals the grain dimensionality due to the influence of thermal energy on magnetic anisotropy.[41,42] The AHE loops under various temperatures are shown in Figure S23 of the Supporting Information. $H_c$ at each temperature of different states are extracted in Figure 3e. As the temperature decreases, $H_c$ exhibits a more rapid increase with a reduction in grain size (from FM-MDL to FM-SD state). This phenomenon is due to the presence of thermal energy, which spontaneously reverses the magnetization of a single domain.[41,43] Consequently, the total magnetization decays as a function of time, which can be described as

$$M = M_0 exp(-\frac{t}{\tau}). \tag{3}$$

The relaxation time, $\tau$, exhibits both temperature and particle size dependency,

$$\tau = \tau_0 exp(\frac{K_u V}{k_B T}), \quad (4)$$

where $\tau_0$ is the inverse frequency factor, $V$ is the particle volume, $K_u$ is the anisotropy constant, $k_B$ is the Boltzmann constant, and $T$ is the temperature.[4] The thermal fluctuation slows down ($\tau$ increases) with the decrease in temperature. Therefore, the observed escalation in temperature dependency from FM-MDL to the SP-SD state aligns with the reduction in grain size as depicted in Figure 3e. At low temperatures, the system stabilizes as $\tau$ becomes comparable to the measuring time ($\tau_m$), hence ferromagnetic properties emerges at temperatures below the blocking temperature ($T_B$).

$$T_B \approx \frac{K_u V}{k_B ln(\tau_m/\tau_0)}. \quad (5)$$

$T_B$ of the SP-SD state is much lower than the other states due to its single-domain characteristic at this state.[41]

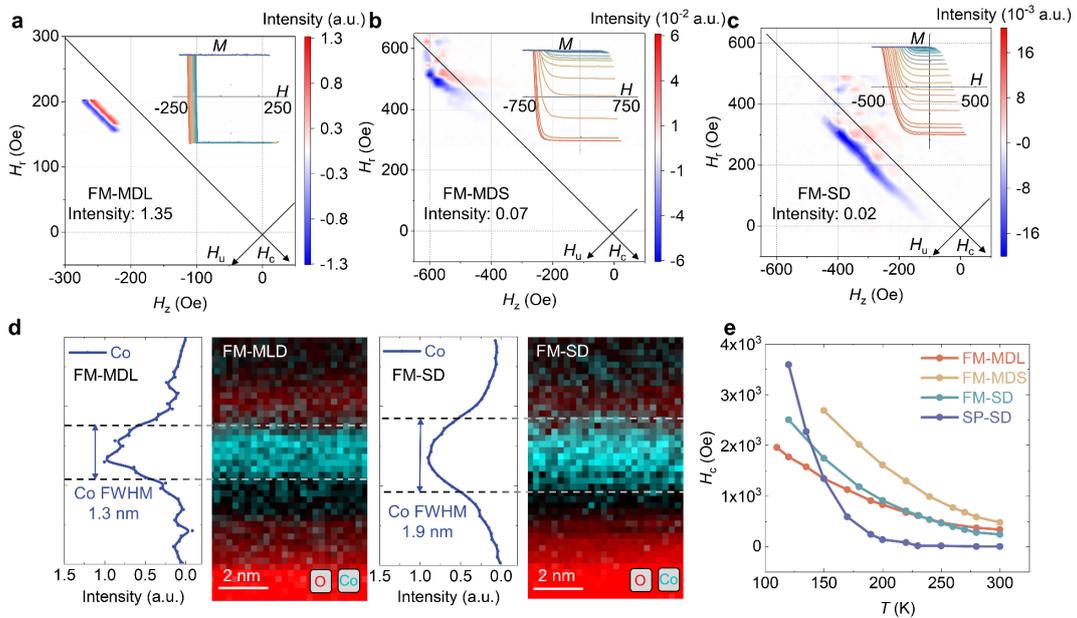

Figure. 3. Elemental redistribution from electrochemical gating. (a-c) FORC results of the device at FM-MDL, FM-MDS, and FM-SD state, respectively. (d) EELS elemental analysis across the Co-Pt interface at FM-MDL state (left panel) and FM-SD state (right panel). In the EELS elemental mappings, the red pixels represent the O element, and the blue pixels represent the Co element. Scale bar: 2 nm. (e) $H_c$ variation as a function of the temperature of different states from 100 to 300 K.

Electrical control of $H_c$ serves as a method to achieve tunable pinning sites for domain wall motion. A polar magneto-optical Kerr effect (MOKE) microscope is applied to visualize the domain pinning effect upon gating. Figure 4a displays the MOKE image of two devices. Within the dashed circle, the ion migration transformed the device to FM-MDL ($H_c$ = 190 Oe) and FM-MDS ($H_c$ = 450

Oe) states, by applying $V_G$ = -2 V for 10 and 100 s, respectively. The region outside the circle remains the virgin state ($H_c$ = 160 Oe). Due to the non-volatile nature, the modulation persists after the removal of gate voltages. For each device, a perpendicular magnetic field, $H_z$, slightly lower than the $H_c$ of the FM-MDL or FM-MDS region, is initially applied to observe domain nucleation. Subsequently, multiple magnetic field pulses of 10 ms are applied to observe the pumped domain wall motion. As illustrated in Figure 4a, in the FM-MDL state under $H_z$ = 160 Oe, a domain immediately nucleates, propelled by magnetic field pulses and moves across the Hall bar area within 10 ms. In contrast, in the FM-MDS state under $H_z$ = 400 Oe, domain nucleation initially occurs in the virgin region, and slowly moves across the gated region. Eventually, the domain wall passes the Hall bar area in 4 s. More MOKE images of domain wall motion are provided in Figure S9 of the Supporting Information.

To further investigate the influence of grain size on domain nucleation and motion, Figure 4b displays the time-dependent MOKE intensity under constant magnetic fields. The magnetization in the FM-MDL region immediately switches at a minimum nucleation field of 150 Oe, while in the FM-MDS region, it took 5 s for the magnetization to switch at a minimum nucleation field of 280 Oe. We then evaluate the domain dynamics at different amplitudes of magnetic fields in Figure 4c. We observe that domain nucleation takes longer in the FM-MDL state (depending on the pulse amplitude), but the magnetization switches immediately. In contrast, domain wall motion across the gated area is slower at FM-MDS and FM-SD states. As $H_z$ increases, the domain wall moves faster for grains of all sizes. Further, we calculate the domain motion velocity as a function of the nucleation field in Figure 4d, indicating that the reduction in grain dimensionality significantly suppresses domain wall motion by two orders of magnitude.

To harness this gating-controlled domain pinning effect, we present a device featuring an array of gating regions to pin the domain wall at specific locations (Figure 4e). Upon the application of $V_G$, the formation of EDL only facilitates $O^{2-}$ ion migration in the unprotected area, resulting in grain refinement, while leaving the large grains intact at the photoresist-covered areas (Figure S8, Supporting Information). We exploit this approach to build a multilevel magnetic state device. Figure 4f schematically illustrates the magnetic field-driven domain motion across the gated area. The refined grains at the gated area pin the domain motion, causing the magnetization reversal at a higher magnetic field than the protected area. In Figure 4g, we gated six strip regions in the channel area at $V_G$ = -2 V for 10 s. Under an external magnetic field, the magnetization at the protected region reverses first, followed by a reversal at the gated area. Eight distinct levels are observed for the sweep of magnetic fields of opposite directions.

Furthermore, we explore the impact of this pinning effect on current-driven magnetization switching. A charge current in the heavy-metal layer can be converted into a perpendicular spin current, leading to a magnetization switching (Figure 4h).[44,45] We observe that reducing grain dimensionality significantly lowers the spin-orbit torque (SOT) efficiency (Figure S29, Supporting

Information), thereby enabling the realization of multilevel current-driven magnetization switching. To assist this SOT switching, a small assistive field is applied in the z-direction to overcome the Joule heating due to low Pt thickness (used in our stack). Figure 4i shows the MOKE signal of the current-driven domain wall motion. The application of $V_G$ = -2 V for 10 s at the two strips leads to the grain refinement. Four levels of magnetization switching are observed as a function of current, with a pulse width of 100 µs. Levels 0 and 3 correspond to the saturation magnetization in opposite directions. The red (blue) curve illustrates positive (negative) magnetization reversal when ramping up (down) the pulse current under a positive (negative) $H_z$. The inset of Figure 4i displays MOKE images at four representative levels. Initially, the magnetization reverses on one side of the Hall bar device under the assisted magnetic field. A higher current is required to switch the magnetization in the gated area with a smaller grain size. To optimize the device for future applications, one could increase the thickness of Pt to take advantage of SOT.

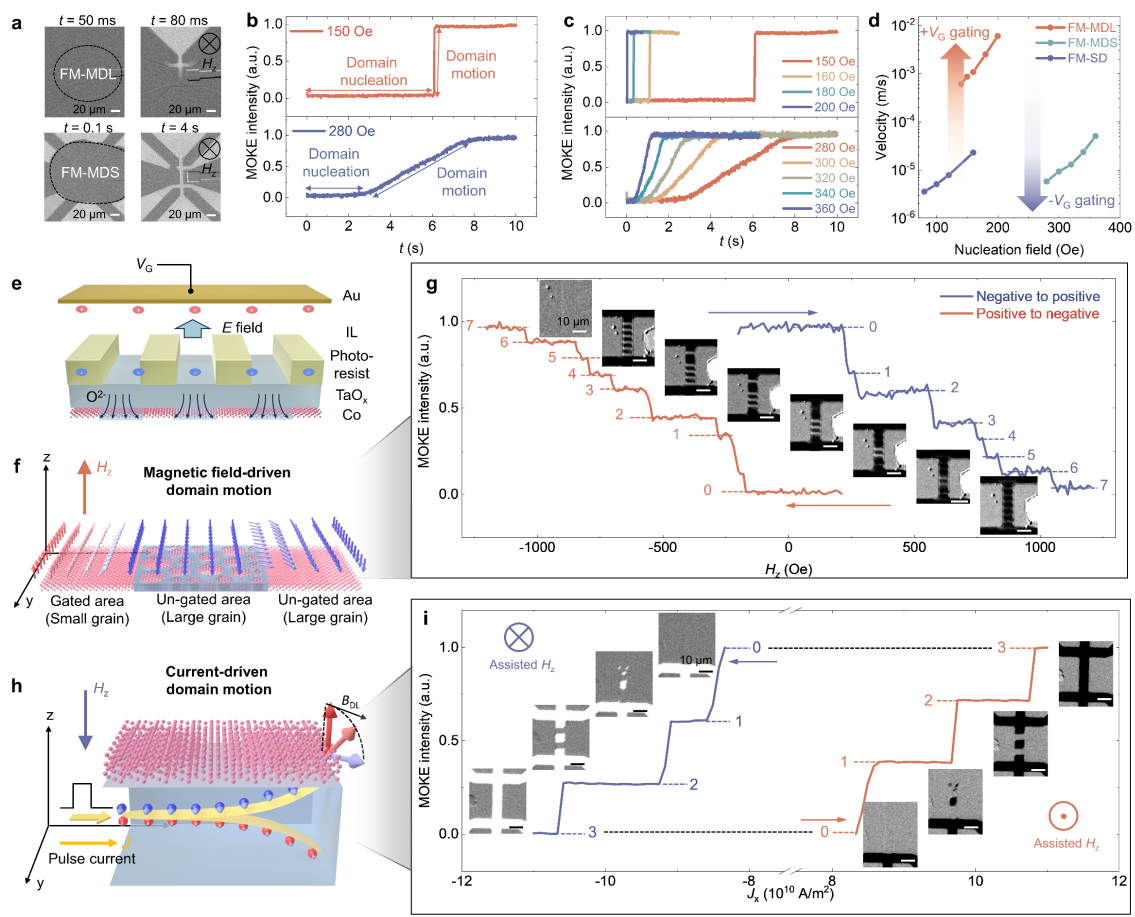

Figure 4. Selective area grain refinement for multilevel magnetization switching. (a) MOKE image of the domain nucleation at FM-MDL and FM-MDS states. The scale bar represents a length of 20 µm. (b) Time-dependent domain nucleation and motion at FM-MDL and FM-MDS states with the minimum magnetic field required for the domain nucleation. The applied field for both states is the minimum field required for nucleation. (c) Time-dependent domain nucleation and motion under various magnetic fields. (d) Domain motion velocity as a function of nucleation field at different

states. (e) Schematic of selected region gating induced local grain refinement. (f) Schematic of the magnetic field-driven domain motion across the gated area. (g) Magnetic field-driven multilevel domain motion. The gated area exerts a pinning effect on domain motion, resulting in sequential magnetization reversal. Scale bar: 10 µm. (h) Schematic of SOT switching. (i) Current-driven multilevel domain motion. With increasing current amplitude, the domain gradually migrates towards the other side of the Hall bar device. The magnetization of the two strips reverses due to the higher SOT requirement. This non-sequential magnetization reversal in our setup could be attributed to the collaboration of SOT and spin transfer torque (STT).

**Conclusions**

In summary, we demonstrate an electrical control magnetic anisotropy via tunable grain dimensionality. The application of $V_G$ at low voltages induces a transition from a ferromagnetic state with large grains to a superparamagnetic state characterized by small grains. This transition is marked by an initial increase in $H_c$ to its maximum value, followed by a gradual decrease to zero. Importantly, the modulation is both non-volatile and reversible, a characteristic of electrochemical gating. Through comprehensive electrical and elemental analysis, the change in magnetic anisotropy is attributed to the grain dimensionality variation induced by $O^{2-}$ ion migration. Furthermore, we observe that the gating-induced small grains can function as pinning sites for domain wall motion. Leveraging this electrically controlled pinning strength, we extend its applicability to magnetic field and current-driven multilevel magnetization switching devices. These findings provide insight into the robust control of magnetic anisotropy through electrical modulation of grain dimensionality in magnetic thin films.

**Methods**

*Device fabrication*

The film was grown on $SiO_2$ substrate using a magnetron sputtering tool. All targets were sourced from AJA International, with a purity of 99.9% or better. The chamber was initially evacuated to $10^{-8}$ Torr before sputtering. We used DC magnetron sputtering mode to deposit Ta and Co, deposition rates of 0.0177, and 0.021 nm/s, respectively. RF magnetron sputtering mode was employed to deposit Pt, with a deposition rate of 0.013 nm/s. For all depositions, the power was set at 50 W in an Ar ambient of 3 mTorr.

Subsequently, Hall-bar devices were patterned using UV lithography, 10 nm was etched down using Ar ion milling technique. The upper $TaO_x$ layer was formed through oxygen plasma treatment. The chamber was first evacuated to $10^{-3}$ Torr before injecting $O_2$ at a flow rate of 20 sccm, resulting in a pressure of $5\times10^{-1}$ Torr. The plasma treatment was conducted at 50 W for 30 s, transforming the ultrathin Ta layer into an amorphous TaO$_x$ layer. Electrodes were patterned using UV photolithography, followed by the deposition of Cr (5 nm)/Au (50 nm) using a thermal evaporator. All the UV lithography work was performed via an Ultraviolet Maskless Lithography machine (TuoTuo Technology (Suzhou) Co., Ltd.). After lift-off in acetone and dried with N$_2$ flow, a drop (10 µL) of ionic liquid *N,N*-diethyl-*N*-(2-methoxyethyl)-*N*-methylammonium bis-(trifluoromethylsulphonyl)-imide (DEME-TFSI from Kanto Chemical) was applied at the center of the Hall-bar device, covering the gate electrode and the channel.

*Electrical measurement*

The room temperature electrical measurements were conducted in a Janis ST300 cryostat equipped with a dipole magnet, with a magnetic field sweeping rate of 10 Oe/s. A vacuum of $10^{-2}$ mbar was maintained to avoid the influence of ambient moisture. For longitudinal and transversal resistance measurements, a current source of 20 µA was applied by a Yokogawa GS200, and voltages were monitored using a Keithley 2000. Meanwhile, Keithley 2400 was used to apply gate voltages. The low-temperature, high magnetic field measurements were performed in an Oxford TeslatronPt cryostat equipped with a superconducting magnet up to 80,000 Oe, with a magnetic field sweeping rate of 30 Oe/s. The spin-orbit torque (SOT) efficiency was obtained by performing harmonic electrical measurement, an AC current was applied between S and D electrodes at a frequency of 17.7 Hz using a Keithley 6221 current source. First and second-harmonic voltages were obtained using two Stanford SR830 Lock-in amplifiers in the transverse direction.

*MOKE Measurement*

A polar MOKE microscope was used to study the domain wall dynamics of electrically gated regions. Before every measurement, the magnetization of samples was saturated by applying a field larger than the saturation field in the perpendicular direction to the sample plane. Subsequently, a magnetic field pulse of amplitude (in the opposite direction of the saturation pulse) less than the coercive field of the sample was applied to nucleate the opposite magnetic domain. We applied magnetic field pulses at 10 ms to observe the magnetic field-driven dynamics of the domain wall. For observing the current-driven dynamics of the domain wall, we applied current pulses of 100 µs in the presence of a small assistive field in the z-direction. A Keithley 6221 source metre was used to apply the current pulses. The assistive field was needed to prevent the joule heating of samples, as a large current amplitude was required to get a sufficient proportion of current to pass through the thin Pt layer.

**Supporting Information**

The Supporting Information includes several sections that provide additional details on the characterization of the deposited film and the dimensions of the device. More information on the reversible and long-term repeatability of both resistance and $H_c$ modulation under gate voltages. Validation of $O^{2-}$ ion migration across the Co/TaO$_x$ interfaces, and investigation of gating influence on SOT efficiency.

**Acknowledgements**

The authors acknowledge the National Research Foundation (NRF) Singapore funding for the CRP21 grant NRF-CRP21-2018-0003. S.Li acknowledges CRP for the research scholarship. S.B. and S.N.P. acknowledge the financial support by the Ministry of Education, Singapore, under its Tier 2 grant MOET2EP50122-0023. X.R.W. acknowledges support from Singapore Ministry of Education under its Academic Research Fund (AcRF) Tier 1 (Grant No. RG82/23), Tier 2 (Grant Nos. MOE-T2EP50120-0006 and MOE-T2EP50220-0005) and Tier 3 (Grant No. MOE2018-T3-1-002), and the Agency for Science, Technology and Research (A*STAR) under its AME IRG grant (Project No. A20E5c0094). X.R.W. and S.N.P. designed and directed this study.

**Competing interests**

X.R.W, S.N.P and S.L are co-inventors on a patent application (Singapore provisional filing number 10202303019Q) related to the research presented in this paper.